\documentclass[runningheads]{llncs}
\usepackage{amsmath}
\usepackage{amsfonts}
\usepackage{cleveref}
\usepackage{graphicx}
\usepackage{multirow}
\usepackage{cite}
\usepackage{floatrow}
\usepackage{acronym}
\usepackage{xurl}
\newfloatcommand{capbtabbox}{table}[][\FBwidth]

\usepackage{blindtext}
\usepackage[numbers]{natbib}
\acrodef{MOOC}[MOOC]{Massive Open Online Course}
\acrodef{MPC}[MPC]{Most Popular Completion}
\acrodef{MS}[MS]{Microservice}
\acrodef{IR}[IR]{Information Retrieval}
\acrodef{QAC}[QAC]{Query Autocomplete}
\acrodef{QS}[QS]{Query Suggestion}
\acrodef{SIR}[SIR]{Social Information Retrieval}

\begin{document}

\title{FullBrain: a Social E-learning Platform}
%
%
\author{
Mirko Biasini\inst{1} \and
Vittorio Carmignani\inst{1} \and
Nicola Ferro\inst{2} \and
Panagiotis Filianos\inst{1} \and
Maria Maistro\inst{3} \and
Giorgio Maria di Nunzio\inst{2}
}
\authorrunning{M. Biasini et al.}
%
\institute{FullBrain, \url{https://fullbrain.org/}\\
\email{\{mirko, vitto, panos\}@fullbrain.org}\and
University of Padua, Padova, Italy\\
\email{\{ferro,dinunzio\}@dei.unipd.it}\and
University of Copenhagen, Copenhagen, Denmark\\
\email{mm@di.ku.dk}
}
%
\maketitle              

\begin{abstract}
We present FullBrain, a social e-learning platform where students share and track their knowledge. FullBrain users can post notes, ask questions and share learning resources in dedicated course and concept spaces. We detail two components of FullBrain: a \ac{SIR} system equipped with query autocomplete and query autosuggestion, and a Leaderboard module to improve user experience.
We analyzed the day-to-day users' usage of the \ac{SIR} system, measuring a time-to-complete a request below $0.11$s, matching or exceeding our UX targets. Moreover, we performed stress tests which lead the way for more detailed analysis.
Through a preliminary user study and log data analysis, we observe that $97\%$ of the users' activity is directed to the top $4$ positions in the leaderboard.
\keywords{e-learning  \and social networks \and social information retrieval \and gamification \and leaderboards}
\end{abstract}

\section{Introduction}
    
E-learning platforms have been improving the quality of education in the digital sphere, with an increasing popularity over the past two decades.
Nevertheless, the main approach has been to port the offline experience to its virtual analogue, as in the case of \ac{MOOC} platforms, which offer video lectures, note taking, quizzes, forums, etc. 
Well known examples of \ac{MOOC} platforms are edX~\cite{Edx2020}, Udemy~\cite{Udemy} and Coursera~\cite{Coursera2020}. 
However, \ac{MOOC}s struggle to keep users engaged: approximately $88\%$ of learners \emph{do not return} after a first year of usage and on average only $8\%$ of students per cohort are retained~\cite{themoocpivot}. \newline
Moreover, we believe \ac{MOOC} platforms to be structurally lacking. 
By mimicking the physical education experience, they miss the possibility of providing the novel experience that the digital medium can offer. A multitude of digital learning features has been proposed, such as personalization, gamification and social, connectivist learning, however the attempts to align these elements have been sparse. In this context, we identify two main challenges.\newline
First, the social component is a fundamental part of the learning process~\cite{socialla, gil, liu, feedmeburke}, and when this is not provided by e-learning platforms, it is usually supplemented, in official university courses, by different services, as Slack~\cite{Slack2019}, Teams~\cite{Microsoft2020}, Piazza~\cite{Piazza2017}, Moodle~\cite{Moodle2018}, etc. However, many students consider these solutions inadequate and, when possible, they prefer to directly contact a peer they personally know and trust. Between each other, they use makeshift solutions like Facebook Groups or Messenger~\cite{Messenger} and WhatsApp~\cite{WhatsApp} group chats. Students there share the same context and can immediately assist on queries with understanding of the course requirements. The above, however, force users to combine a multitude of services to achieve all their goals.
Second, a single centralized and reliable source of learning material is not available for online learners. Students often search and access to different sources through Web search engines, however they struggle in determining the quality of each resource, especially when they do not have enough knowledge to critically review them. The results are often ordered following an opaque process. The above require extended time to review learning resources before usage.\newline
The necessity of social e-learning platforms able to address the above challenges has became evident during the COVID-19 pandemic, when there has been a large scale transfer of learning activities to online spaces. 
According to Unesco~\cite{UNESCO2020} over $90\%$ of the world students are confined at home. Providing alternative learning solutions has become the top priority.\newline
We present \emph{FullBrain} a social e-learning platform which addresses these challenges. FullBrain is framed in the context of connectivism theory~\cite{goldieconnectivism,siemensconnectivism}, which represents learning as a network phenomenon influenced by technology and socialization.
FullBrain offers open, dedicated spaces corresponding to university courses and concepts. It allows students to interact with these spaces, to post thoughts, ask questions and share, rate and comment online learning resources, tagging them with instructional metadata. These learning resources are organized in a crowdsourced ontological structure forming a reviewed \emph{digital library}. \newline 
In this paper, we first present FullBrain's features and then focus on two fundamental components of its architecture: \acf{SIR} System, that powers our search algorithms, and a user leaderboard, that encourages social connectivity in concept and course spaces.
This paper is organised as follows:~\Cref{sec:relatedwork} details the related work;~\Cref{sec:thesystem} introduces FullBrain's design, features and system's architecture;~\Cref{sec:socialinformationretrievalsytem} and ~\Cref{sec:leaderboard} present the \ac{SIR} and the leaderboard components respectively;~\Cref{sec:prototype} showcases the prototyped solutions, which are evaluated in~\Cref{sec:evaluation}. Finally,~\Cref{sec:conclusionsfuturework} includes conclusions and future work.
\section{Related Work} \label{sec:relatedwork}
    
    FullBrain belongs to the category of social e-learning platforms. 
    Such efforts rarely have been part of literature  
    and FullBrain stemmed from a few research papers: 
    Gil and Martin-Bautista~\cite{gil} propose a connectivist, ontological knowledge support system with a focus on collectivist knowledge building; Liu et al.~\cite{liu,liutaskbased} propose systems for collaborative relevance assessment with tasks as context; and Chen and Su~\cite{chenandsu} propose SAP, an ontology-based social learning platform with learning paths.
    
    However none of these approaches have been adopted in a larger scale. 
    edX~\cite{Edx2020}, Coursera~\cite{Coursera2020} and Udemy~\cite{Udemy} are \ac{MOOC} platforms that provide content, but no immediate social networking tools. 
    In contrast, learning platforms that offer social learning, as Golden~\cite{Golden} and Expii~\cite{Expii}, have a wiki approach to knowledge. 
    Differently, FullBrain fully integrates the learning and social components in a single platform.
    
        In addition to the social networking features, FullBrain offers a \ac{SIR} System~\cite{Kirsch2006}, which ranks results using both topical and user similarity. The user similarity is calculated based on the distance in the social graph between the searcher and the results~\cite{Kirchhoff2008, Vieira2007, Lampe2006}. We propose an architecture of an IRS specifically designed in the context of Online Social Networks \cite{Boyd2007} able to handle different entities, rather than only users (k-$n$ partite relationships \cite{Bouadjenek2016a}). To the best of our knowledge such architecture is novel.
        
        Finally, FullBrain exploits gamification approaches through a leaderborad. Gamification has proven to be beneficial to increase user engagement in numerous contexts~\cite{costa2013a, landers2017a, ortiz-rojas2019a, landers2014a}. Leaderboards are a workhorse of gamification~\cite{leung2019a} and are considered one of the key elements in game design~\cite{hickman2010a}. According to Zichermann~\cite{zichermann2011}, leaderboards can follow two approaches: absolute and relative. Upon these, leaderboards emphasize continuous performance, status reporting, comparison, socialization, and competition~\cite{christy2014a, butler2013a}.

\section{FullBrain System}
\label{sec:thesystem}

In this section we present the elements that comprise FullBrain's user interface and system. We start with the description of the main features in~\Cref{subsec:mainactors}, the underlying social graph in \Cref{subsec:socialgraph} and the overall architecture in \Cref{subsec:architecture}.

\subsection{Features}\label{subsec:mainactors}
Six entities represent the core of FullBrain's offering: Posts, Streams, Sources, Courses, Concepts and User Profiles. We detail them next.

As in any regular Social Network, the users are able to create \textbf{posts} to share ideas. Also, users can post anonymously and specify if it is a question or a note. In the foreseeable future, they will be able to mark a question as `resolved' if the issue was addressed. Posts are feature-heavy: 
users can like, comment and share any post they have access to.
    
\textbf{Streams} 
refer to collections of consecutive posts in a specific space: in the context of a course, concept or user profile page.
The user home page is an example of stream, where users receive posts aggregated from streams of the entities they have \textit{included} (followed).

    An operational definition of \textbf{source} is: \textit{a resource to learn one or multiple subjects which can be freely accessed online}.
    Simply, sources are URL links pointing to a destination website which provide learning content. Sources are shared by members of the FullBrain community. These resources combined with the user who shared them have a set of metadata that form the source card:
    \begin{itemize}
    	\item \textit{Name}, \textit{affiliation} and \textit{level} of the user who shared the resource. The level is an indicator of the user's reputation in the platform. It depends on, for example, user's interactions or obtained reviews on shared resources. This will be further addressed in~\Cref{sec:leaderboard};
    	\item \textit{Learning styles}, which can be: watching, reading, practise/exercise, listening, group activity or tutorial; 
    	\item \textit{Title} which highlights the content of the source;
    	\item \textit{Instructions} on how to traverse the source. For instance, if the source is a video, the instructions may be: ``watch from minute 2:03 to 5:40 to learn concept X'';
    	\item \textit{Prerequisites}: concepts that users have to know in advance to understand the learning source. Clicking into a prerequisite results in opening the related concept page;
    	\item Social stats: \textit{star reviews} and \textit{comments} allow to express an opinion on the source relevance for future viewers.
    \end{itemize} 
    
    We refer to \textbf{course} as an academic course associated with its respective Institution. 
    Examples can be `Machine Learning' and `Database Systems' at the University of Padua (Unipd) or `Business Management' at the Copenhagen Business School (CBS). 
    Each course has a dedicated space with group features:
    users can join and post thoughts and questions. Moreover, students can share learning resources 
    helpful to pass the course/exam. 
    The course's sources slightly differ from the ones in concepts: they do not have prerequisites, but tags, which allow to label
    the material based on the purpose: \#lecture1, \#exams or \#exercises.

An operational definition of \textbf{concepts} can be as follows: \textit{the smallest unit of knowledge, idea or process, which can not be divided further without becoming an other concept or losing utility for the task at hand}.

    It is important to note that this is a soft definition, a guideline which is open for interpretation by FullBrain's members. Every FullBrain member can create a concept space, which allows post and resource sharing. In that respect interactions are very similar to Course spaces. Concepts are linked through learning resource prerequisites, forming crowdsourced ontologies. Moderating the validity of these ontologies is expected to be a social endeavor as well. A user experience workflow is not proposed currently and is matter of future research.
    
    Each user has a \textbf{user profile} page, which includes the user name, institutional affiliation, description and picture. In addition it shows the list of includers (followers) and included users (followed users), the lists of the courses and concepts the user has included and a stream of their posts.
            
\subsection{Social Graph} \label{subsec:socialgraph}
    A social network can be seen as a graph~\cite{Nasution2016, Nasution2017, Elveny2018}.
    Indeed, Elveny et al.~\cite{Elveny2018} present the users as social actors which interact with each other creating relationships. 
    In FullBrain, the relationships between users are: asymmetric, explicit, non-confirmed, regular and unsigned~\cite{Guy2018,Guy2010}.
    
\begin{figure}[t]
        \centering
        \includegraphics[width=0.7\textwidth]{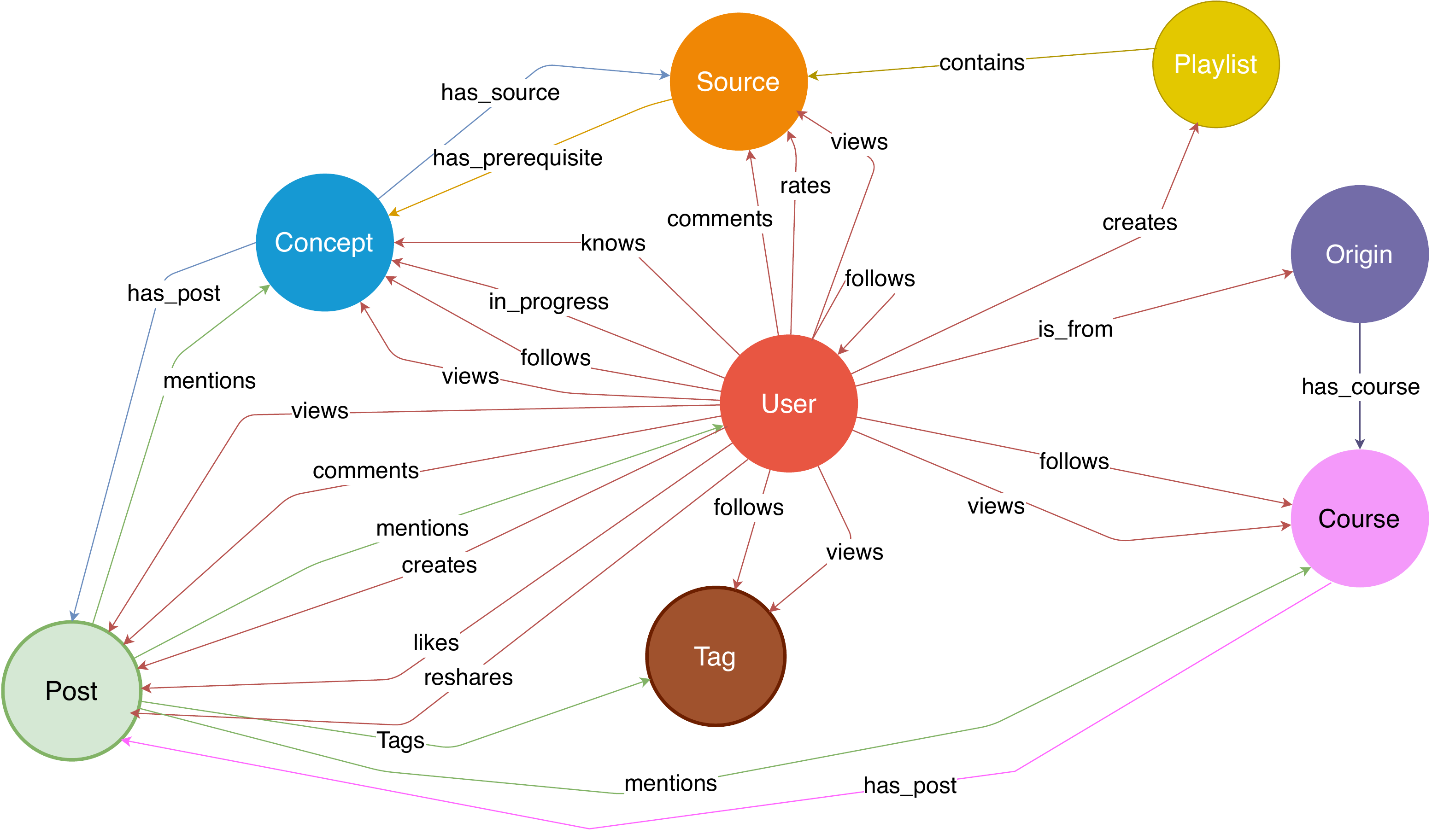}
        \caption{FullBrain Social Graph with entities as nodes and related relationships as edges.} \label{fig:DataDiagram}
    \end{figure}

    The FullBrain social graph, shown in~\cref{fig:DataDiagram}, can be seen as an extended social graph with the $k$-partite definition~\cite{Bouadjenek2016a}, where nodes can be of different types. We recognize the main \emph{entities} in \Cref{subsec:mainactors} as nodes. In addition, we find \emph{Origins} (institutions), \emph{Tags} (labels for learning material and posts) and \emph{Playlists} (public or private collections of favorite sources created by users). As edges, we can see all the possible relationships between the various entities (nodes).
    
\subsection{Architecture} \label{subsec:architecture}
\begin{figure}[t]
        \centering
        \includegraphics[width=0.7\textwidth]{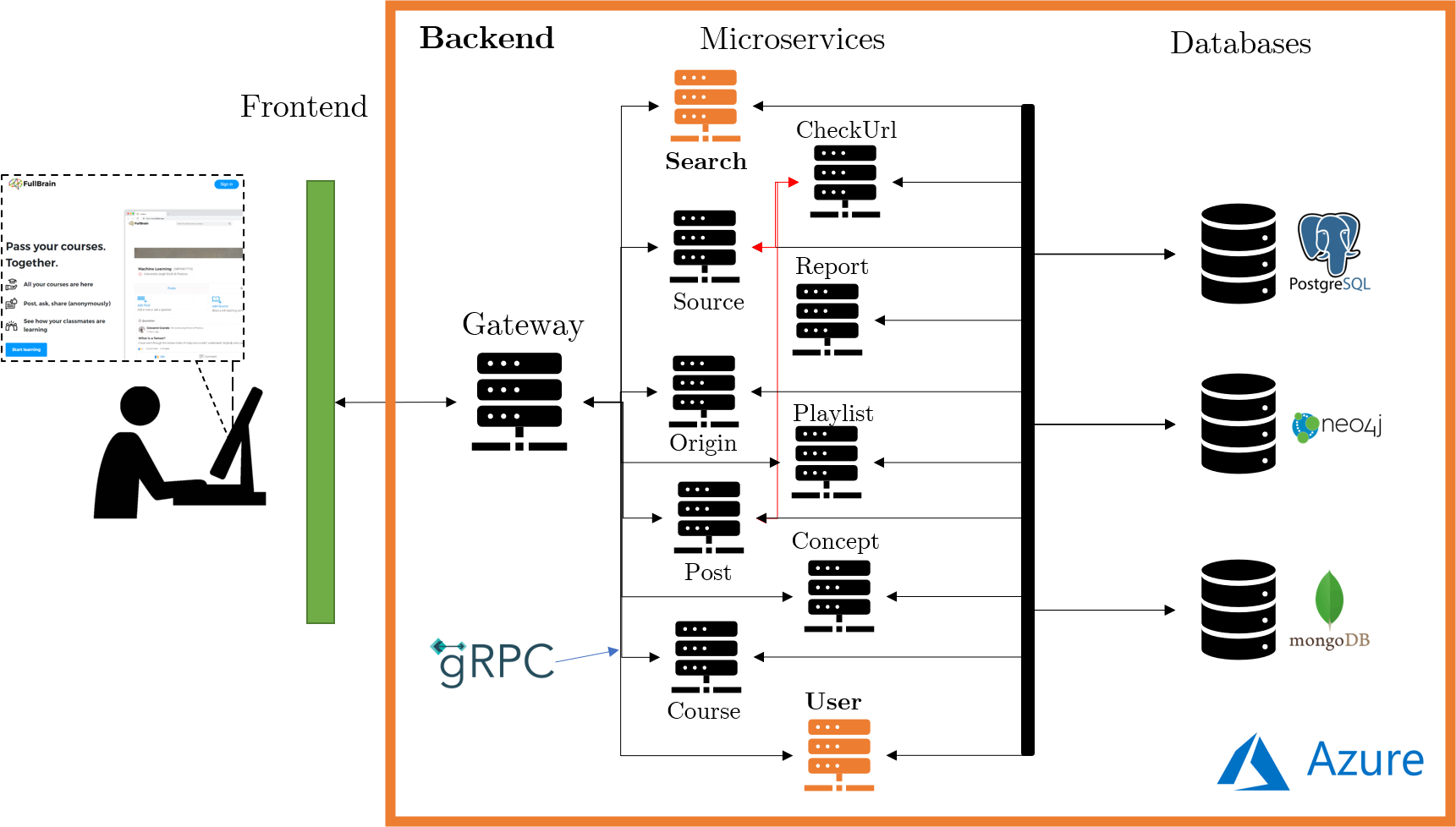}
        \caption{FullBrain Architecture. In orange the Microservices (MS)s that contain the functions we describe in \Cref{sec:socialinformationretrievalsytem} and \Cref{sec:leaderboard}. } \label{fig:architecture}
    \end{figure}

    \Cref{fig:architecture} displays FullBrain's architecture.
    All virtual machines and databases are hosted in Azure~\cite{Microsoft2019}. Cloud services ensure scalability and continuity. We split FullBrain's architecture into three main parts: \emph{front-end}, \emph{back-end} and \emph{databases}.
    
    The front-end is the layer in direct contact with the user. When a user navigates through the FullBrain platform, it creates a series of requests. Those are forwarded from the front-end browser client to the back-end server. The back-end is in charge of handling those requests, and return back the expected results.
    
    Multiple microservices compose the back-end~\cite{Mazzara2017}. To each \ac{MS} we assign specific functionalities. \ac{MS}s can communicate using gRPC~\cite{Google} (a language independent protocol) using Protobuf serialization. 
    Such architecture gives the possibility of developing each of the \ac{MS}s in a different programming language to maximize the performance based on peculiarities and available libraries.
    
    Finally, as we can see in~\cref{fig:architecture}, each \ac{MS} is connected to different databases, depending on their operations/calls. \textit{PostgreSQL}~\cite{PostgreSQL2013} maintains most of FullBrain's entities' fields data, and it ensures integrity between them. 
    \textit{Neo4j}~\cite{Neo4j2020}, a graph database, manages the different entity relationships seen in~\Cref{fig:DataDiagram}. \textit{MongoDB}~\cite{MongoDB2020} is mostly utilised for logging of user's activities and requests.

\section{\acf{SIR} System}\label{sec:socialinformationretrievalsytem}
Carmignani~\cite{vittoThesis} describes FullBrain's \ac{SIR} system, which resides in the Search \ac{MS} (\cref{fig:architecture}). It offers three functionalities: \emph{Search}, \emph{\ac{QAC}} and \emph{\ac{QS}}. 
Search and \ac{QAC} estimate relevance by computing the similarity between the query, the entities and the user, and then order the results accordingly to such value~\cite{Levene2011}.

    The overall similarity $S$~\cite{Kirchhoff2008} for a query $q$ with respect to an entity $e$ and a user $u$ (searcher) is calculated as follow:
    \begin{equation}
        S(u,q,e)=\frac{\alpha S_T(q,e) + \beta S_U(u,e)}{\alpha + \beta}\in[0,1]\; \; \;  \alpha, \beta \in \mathbb{R}^+
        \label{eq:overall_similarity}
    \end{equation}
    where $S_T$ and $S_U$ are, respectively, \textit{topical} and \textit{user} similarity. We choose $\alpha=\beta=1$ to give the same importance to the topical and user components. 
    
    The user similarity $S_U(u,e)$ in Equation~\eqref{eq:overall_similarity} is computed as follows:
    \begin{equation}
        \label{eq:userSim}
        S_U(u,e) = 1 - \dfrac{d(u,e)}{\max d}\in[0,1]
    \end{equation}
    where $d$ is the length of the shortest path between $u$ and $e$ in the social graph (\Cref{subsec:socialgraph}).
    
    For efficiency reasons, we calculate an approximation of $d$ using a modified version of the landmark embedding method~\cite{Hotz2019, Vieira2007, Tretyakov2011, Boyd2007, Potamias2009, Sommer2014, Madkour2017}. According to Lampe et al.~\cite{Lampe2006} ``people are more likely to search for other users who are part of their real life friends-network, rather than other members''. Thus, we modify the standard algorithm by not saving distances greater than 3 (assuming them as infinite) as suggested by Vieira et. al.~\cite{Vieira2007}, since entities at such distance would not be relevant for the user. This assumption has saved, in our case, more than $80\%$ of landmarks-nodes stored distances.
    
    The computation of the topical similarity $S_T(q,e)$ in Equation~\eqref{eq:overall_similarity} differs for search and \ac{QAC}. We denote it by $S_{Search}$ and $S_{QAC}$ respectively.
    The search function results include: concepts, users, courses, sources and posts. 
    $S_{Search}$, is computed as a combination of partial~\cite{Ji2009, Chaudhuri2009} and exact match as follows:
    \begin{equation}
        \label{eq:topSimSearch}
        S_{Search}(q,e)=\frac{S_{Partial}(q,e) + S_{Exact}(q,e)}{2}\in[0,1]
    \end{equation}
    For partial matching $S_{Partial}(q,e)$, we employ 
    $q$-grams~\cite{Hall1980, Ukkonen1992, Gravano2001, Navarro2005, Kim2007, Li2013}.
    Given the query $q$ and an entity name $n_e$ (e.g.~user or concept name), we apply the same text operations on $q$ and $n_e$, retrieving their respective set of $q$-grams, $G_{q}$ and $G_{n_e}$:
    \begin{equation}
        \label{eq:topSimSearch}
        S_{Partial}(q,e) = \dfrac{|G_{q} \cap G_{n_e}|}{|G_{q} \cup G_{n_e}|}\in[0,1]
    \end{equation}
    
    For exact matching $S_{Exact}(q,e)$, we use the Vector Space Model~\cite{Levene2011} and calculate $S_{Exact}(q,e)$ as the cosine similarity of the vector representations $\vec{q}$ and $\vec{e}$. Specifically for the latter, we use the union of $e$'s fields, like name and description (giving weights according to the fields' importance):
    \begin{equation}
        \label{eq:topSimSearch}
        S_{Exact}(q,e) = cos(\vec{q}, \vec{e}_{union})\in[0,1]
    \end{equation}

    FullBrain's \ac{QAC} 
    makes possible to retrieve a Qlist~\cite{Smith2017} of entities (users, concepts and courses) that are relevant for the user while typing the query itself~\cite{Smith2017}. 
    As demonstrated in a variety of studies~\cite{White2007, Cai2016, Zhang2015, Tahery2020}, such features improve user satisfaction.
    The elements that populate the Qlist are ranked with the overall similarity in Equation~\eqref{eq:overall_similarity}, where the user similarity is computed as in Equation~\eqref{eq:userSim} and the topical similarity is $S_{QAC}(q,e)=S_{Partial}(q,e)$. We use only partial matching to not compromise the speed of retrieval required by QAC. 
    
    Finally, we describe the \acf{QS} component. 
    We refer to \ac{QS}s as the elements which populate the QList before typing any character into the query input form. \ac{QS}s are either past queries (input for search function) or actual entities (direct link to the related web page). Such elements are retrieved based on the user's search history (more recent would result in a higher rank) without any similarity computation. We also suggest trending queries and entities. Those are retrieved using a modified version of the \ac{MPC} model~\cite{Hu2018, Cai2014, Shokouhi2013}; where a time filter is applied following the finding of~\cite{Tahery2020, Wang2017}, which have shown that queries' popularity is heavily affected by the time.

\section{Leaderboard} \label{sec:leaderboard}
    As previously mentioned, gamification can positively impact users' interactions and contributions. Thus, we designed and developed a leaderboard aiming to increase learners' engagement in FullBrain. Our platform offers a multitude of actions, which can be performed in different contexts, such as courses and concepts. Combining different sets of actions and goals, a variety of leaderboards can be generated to address different motivational areas. Upon these, leaderboards provide learners with elements for reflection, comparison, motivation, socialization, and competition \cite{christy2014a, butler2013a}.

    \paragraph{Points system and mechanics}
        To be created, leaderboards require a point system. The point system is a mechanism defining a set of criteria, which are used to assign points to the users. Through these points, a score is computed. Finally, users are ranked based on their final score.
        
        As suggested by Jia et. al \cite{jia2017a}, leaderboards need to provide a sense of fairness to the users. To generate a points system as fair as possible, we categorized all available actions based on two parameters: effort and value. The former assesses the effort required to perform the action, inside and outside the platform. The latter is an evaluation of the value the performed action provides to the platform.
        
        Among all the possible actions available in FullBrain, 11 were selected to build the points system. Precisely, \textit{add}, \textit{share}, \textit{rate}, \textit{comment}, \textit{upvote comment} a source, \textit{add}, \textit{share}, \textit{like}, \textit{comment}, \textit{upvote comment} a post, and \textit{include user/ course/concept}. Several users were interviewed to assign an effort and a value score to each action. Scores were given between 1 and 5. The higher the score, the higher the effort/value. Next, these were averaged out to generate the final points. As a result, the final points system was built as shown in \cref{tab:leaderboardspointssystem}. Every time a user performs one of these actions, the corresponding points are assigned to the user.
        
        \begin{table}[t]
            \caption{Leaderboard points system actions' scores}
            \centering
            \resizebox{\textwidth}{!}{%
            \begin{tabular}{|c|c|c|c|c|c|c|c|c|c|c|c|}
            \hline
            \textbf{Action} & \textbf{Add} & \textbf{Share} & \textbf{Rate} & \textbf{Comment} & \textbf{\begin{tabular}[c]{@{}c@{}}Up Vote\\ Comment\end{tabular}} & \textbf{Add} & \textbf{Share} & \textbf{Rate} & \textbf{Comment} & \textbf{\begin{tabular}[c]{@{}c@{}}Up Vote\\ Comment\end{tabular}} & \textbf{Include} \\ \hline
            \textbf{Entity} & \multicolumn{5}{c|}{\textbf{Source}} & \multicolumn{5}{c|}{\textbf{Post}} & \multicolumn{1}{l|}{\textbf{User Cou. Con .}} \\ \hline
            \textbf{Avg Effort} & 4.4 & 3.3 & 1.6 & 2.5 & 1.1 & 2.5 & 1.4 & 1.3 & 2.1 & 1.7 & 1.6 \\ \hline
            \textbf{Avg Value} & 4.7 & 4.5 & 4.7 & 3.9 & 4 & 4.5 & 4.4 & 4.4 & 4.1 & 4 & 3.1 \\ \hline
            \textbf{Sum} & 9.1 & 7.8 & 6.3 & 6.4 & 5.1 & 7.0 & 5.8 & 5.7 & 6.2 & 5.7 & 4.7 \\ \hline
            \end{tabular}
            }
            \label{tab:leaderboardspointssystem}
        \end{table}
        
        As described above, different types of leaderboards can be generated based on actions and contexts. To do so, several activity metadata are stored every time a user performs an action: the user performing the action, the performed action, the location where the action is performed (e.g. concept, course, user), the object undergoing the action (e.g. source, post, concept, etc), the date and time the action is performed, and the corresponding points. Finally, these activities are filtered based on one or more parameters, to generate the desired leaderboard.
        
        To handle delete actions, such as deleting a post, analogous data are stored but with a negative score instead. Deleting actions affect only the user who performs the deleting action. Users who earned points by interacting with the deleted element remained unaltered. This is done to ensure fairness.
    
    \textit{Leaderboard design}
        To be effective, leaderboards need to be designed based on the audience and the context \cite{hamari2014a, codish2014a, jia2017a, leung2019a}. In FullBrain, concepts have a larger and diverse audience compared to courses. Indeed, as courses refer to a specific university, they are generally restricted to those students who are currently taking that course. Conversely, concepts are followed by learners all around the world. To determine the most appropriate design for each scenario, interviews, A/B tests \cite{nielsenab} and thinking-aloud usability evaluations \cite{h2016a} were conducted aiming to determine four main points.
        \begin{itemize}
            \item \textbf{Leaderboard design:} four different designs were tested: an absolute, a relative, and two hybrids. The absolute solution focuses on the top players, whereas the relative is centered on the active user \cite{zichermann2011, jia2017a}. On top of these, we proposed two hybrid designs, which combine both approaches to exploit their advantages. Each design affects different motivational areas like reflection, competition, or socialization.
            \item \textbf{Leaderboard items:} which and how information should be displayed in the leaderboards. Here, focus was directed towards users' scores, to determine whether showing points is more effective or not.
            \item \textbf{Time:} Leaderboard points can be calculated based on user activities included in a time window of the current week, month, semester etc. We should determine which time windows provide valuable information to users.
            \item \textbf{Leaderboard type:} identify which type of leaderboards are more interesting for users. For example, we can have \textit{Top Contributor}, \textit{Top Reviewer} or \textit{Top Influencer} leaderboards. Each ranks users based on points collected from a different set of activities.
        \end{itemize}
        The results of these experiments led to the final designs described in the following section. Finally, a specific leaderboard is generated for each concept and course. Note, the way we store data allows creating general leaderboards, comprising groups of concepts and/or courses. For example, the top contributors of a specific university or of a big subject like Machine Learning. More research has been conducted on that subject by Biasini \cite{mirkoThesis}.

\section{Prototype} \label{sec:prototype}
\begin{figure}
            \centering
            \includegraphics[width=\textwidth]{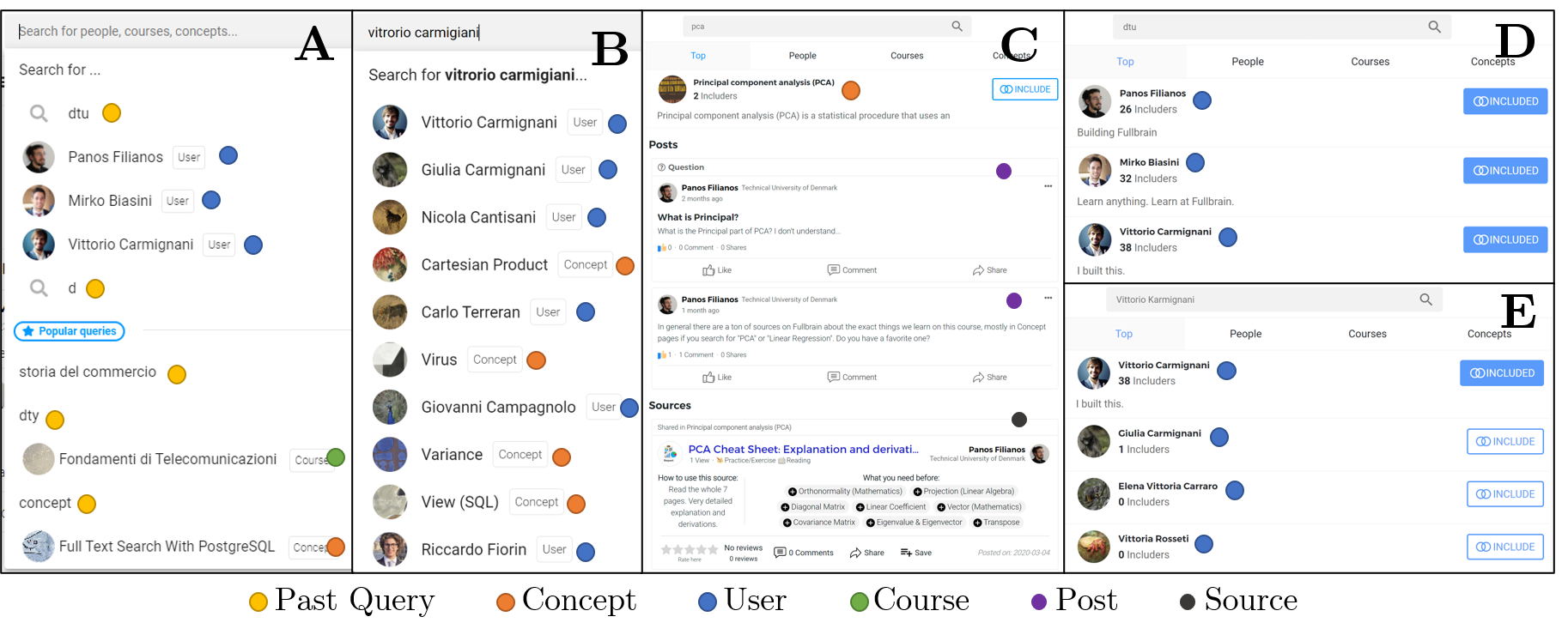}
            \caption{Example of QS (A), QAC (B) and Search Results (C,D,E) for the user ``Vittorio Carmignani''. Moreover, the legend underneath highlights the different entities.} \label{fig:qsAndSearchCrop}
        \end{figure}

    \textbf{Social Information Retrieval System -} \Cref{fig:qsAndSearchCrop} displays the results of using FullBrain's SIR in the FullBrain main search bar. \Cref{fig:qsAndSearchCrop}A shows the QS, where five past queries and entities are suggested together with as many popular ones. \Cref{fig:qsAndSearchCrop}B instead, shows a QAC output and how the use of trigrams permits to retrieve ``Vittorio Carmignani'' despite the input typos. Lastly, \Cref{fig:qsAndSearchCrop}C,D,E show three different example of the search function. In \emph{C}, $q$=``pca'', we can recognize the concept, together with posts and sources. In \emph{D}, $q$=``dtu'' (Technical University of Denmark), the results are retrieved even if not present in the entities'names. This is because we are using additional fields in the search process. Ultimately, in \emph{E}, $q$=``Vittorio Karmignani'', despite the typo we can still retrieve the results thanks to the use of partial matching in the search function. A working example of the \ac{SIR} system is available online\footnote{\url{https://youtu.be/6YOm-xobCzA}}. 
    
%
		
\begin{figure}[t]
            \centering
            \includegraphics[width=\textwidth]{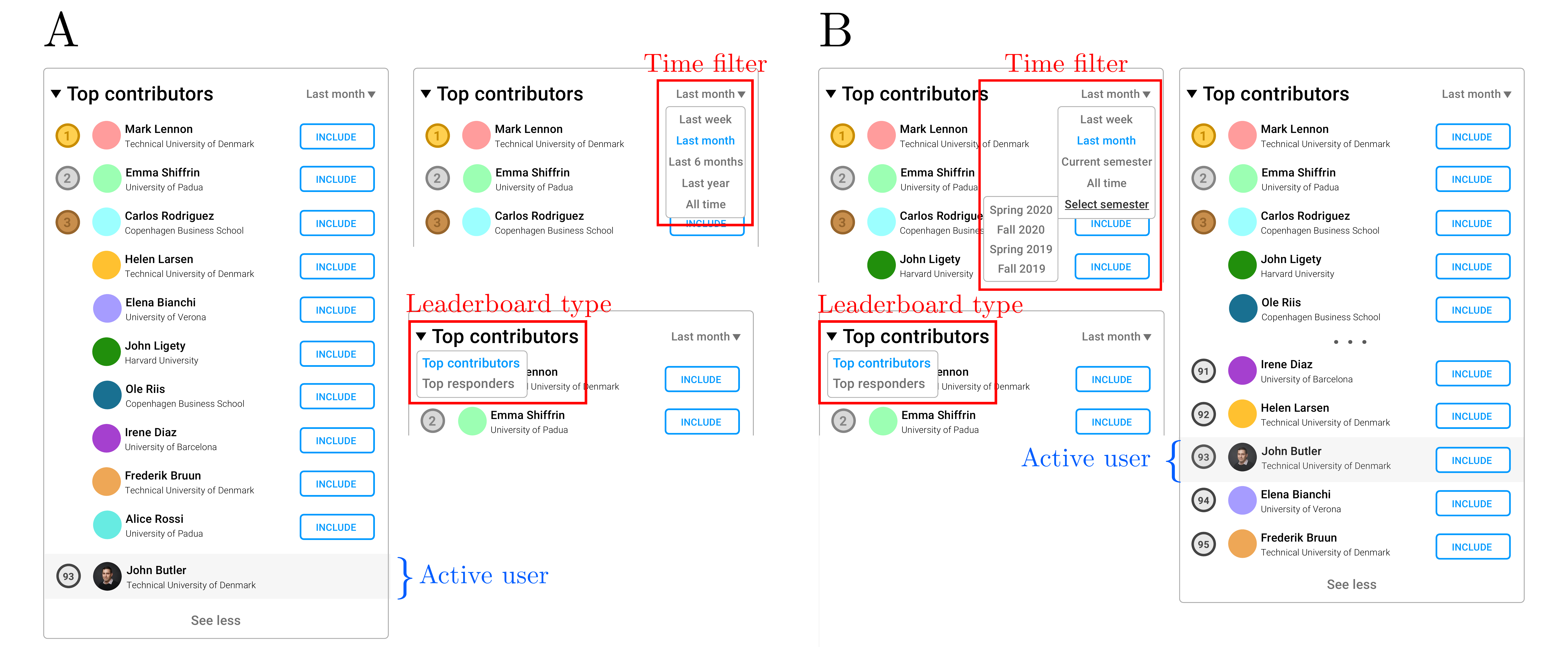}
            \caption{Leaderboard designs for Concept and Course spaces. (A) Hybrid-Absolute design for concepts. (B) Hybrid 50-50 design for courses.}
            \label{fig:leaderboardfinalconcept}
        \end{figure}

    \textbf{Leaderboard -}
        We propose two different hybrid solutions as a result of user interviews, A/B tests and 
        usability evaluations: \textit{Hybrid-Absolute Design} and \textit{Hybrid 50-50 Design} (\cref{fig:leaderboardfinalconcept}).
        We use the hybrid-Absolute design for concepts: an absolute leaderboard shows the top $10$ users and the active user is always shown at the bottom of the leaderboard, regardless of his/her ranking. 
        For courses we use the hybrid 50-50 design, a novel approach combining an absolute leaderboard, on the top, with a relative leaderboard centered on the active user, on the bottom. 
        
        In both solutions, users can access top users' profiles and universities, as well as include them in their home stream. A time filter allows users to change the time window used to build the leaderboard. In addition, users can change the type of the leaderboard choosing between: Top Contributor, which ranks users accounting for all types of actions, and Top Responder, which ranks users accounting just for the frequency of their answers and the quality of their comments. Both solutions are partially implemented in the platform and we are currently working to complete them.
    
\section{Evaluation and Results} \label{sec:evaluation}
\begin{figure}[t]
      \includegraphics[width=0.45\textwidth]{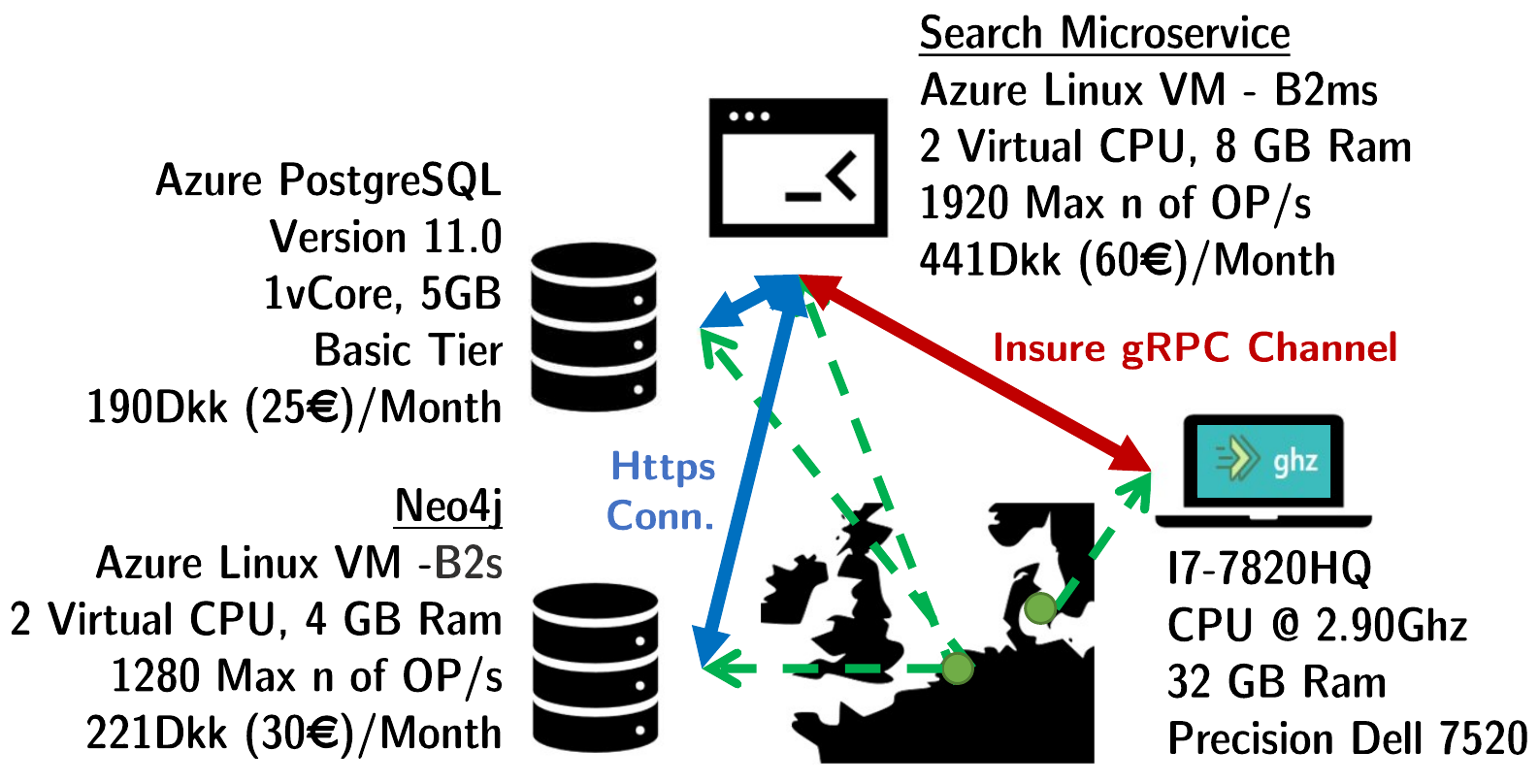}
      \caption{Setup for Stress Tests and components specs.}%
      \label{fig:setupEx}
\end{figure}

    \textbf{Social Information Retrieval System -} Carmignani \cite{vittoThesis} focused on the efficiency of FullBrain's IRS. Based on J. Nielsen's studies \cite{Nielsen1993}, he set that the time to complete a \textbf{QS}, \textbf{QAC} and \textbf{Search} request should be, respectively, \textbf{0.1s} (instantly), in \textbf{[0.1-1]s }(small delay without interrupting user's flow of thought) and \textbf{[1-1.5]s} (it is not an interactive function).  Hence, we have logged the time to complete a QS, QAC and Search requests on FullBrain from all user clients using the platform for 3 weeks. The results are shown in the ``Day-to-day'' usage columns in \cref{tab:effiSir}.  We can see that \textbf{all three functions were within the bounds set}. Therefore, we performed stress tests to test the limits of the architecture. Particularly we made 1000/$n$ requests from $n$ different users in parallel (with $n\in\{1,2,4,8,16,32,64\}$) using \emph{ghz}\cite{Ghz}. The test setup is shown in \cref{fig:setupEx}. The functions were applied on a dataset having a total of 5724 entities and 21512 relationships. The results are shown in the stress test columns in \cref{tab:effiSir}. All functions show a similar performance patterns: they all break the set bounds after $n=4$. In contrast, it is important to remark here that, by using the website, it is not possible to reproduce this kind of loaded environment (with only these number of users): there would be a delay between two users' requests due to the UI and the physical action of submitting them. Although this delay can be minimal, it can be still valuable since in normal conditions a request is answered in 0.1s. In contrast, we know that this delay is zero in case of stress test (since the requests are programmatically made). Thus, we believe that the system can support far more than 64 parallel users using the platform (admitting that the tests machine utilized are not suitable for a production environment). Indeed, between user request submission and its answer computation, there is a whole structure of microservices and communications that takes place.  If the microservices are the problem, then we could create clones and balance the load of requests. If the problem is the computation, we need to know if it is due to physical limits (machine not performant enough) or the actual implementation. Thus, further analysis would be required to understand exactly in which step the bottleneck is of the overall process. 

\begin{table}
\caption{Efficiency Study on FullBrain's SIR based on day-to-day usage on the platform and conducted stress tests.}
\begin{tabular}{c|c|c|c|c|c|c|c|c|c|c|}
    \cline{2-11}
     &
      \multicolumn{2}{c|}{\textbf{\begin{tabular}[c]{@{}c@{}}Day-to-Day\end{tabular}}} &
      \multicolumn{8}{c|}{\textbf{Stress Tests}} \\ \cline{2-11} 
    \multicolumn{1}{l|}{} &
      \multirow{2}{*}{\textbf{Count}} &
      \multirow{2}{*}{\textbf{\begin{tabular}[c]{@{}c@{}}Avg\\Time\end{tabular}}} &
      \multirow{2}{*}{\textbf{Count}} &
      \multicolumn{7}{c|}{\textbf{\begin{tabular}[c]{@{}c@{}}Avg Time (s) for a Request with\\$n$ concurrent users\end{tabular}}} \\ \cline{5-11} 
     &
       &
       &
       &
      \begin{tabular}[c]{@{}c@{}}1\end{tabular} &
      \begin{tabular}[c]{@{}c@{}}2\end{tabular} &
      \begin{tabular}[c]{@{}c@{}}4\end{tabular} &
      \begin{tabular}[c]{@{}c@{}}8\end{tabular} &
      \begin{tabular}[c]{@{}c@{}}16\end{tabular} &
      \begin{tabular}[c]{@{}c@{}}32\end{tabular} &
      \begin{tabular}[c]{@{}c@{}}64\end{tabular} \\ \hline
    \multicolumn{1}{|c|}{\textbf{QS}} &
      8743 &
      0.06s &
      1000 &
      0.039 &
      0.078 &
      0.159 &
      0.3444 &
      0.614 &
      1.191 &
      2.390 \\ \hline
    \multicolumn{1}{|c|}{\textbf{QAC}} &
      7599 &
      0.11s &
      1000 &
      0.506 &
      1.323 &
      0.920 &
      1.912 &
      4.769 &
      7.611 &
      15.029 \\ \hline
    \multicolumn{1}{|c|}{\textbf{Search}} &
      512 &
      0.10s &
      1000 &
      0.234 &
      0.555 &
      0.594 &
      1.751 &
      2.135 &
      1.052 &
      7.754 \\ \hline
    \end{tabular}
    \label{tab:effiSir}
    \end{table}
    

    \textbf{Leaderboard -} During the experiments conducted to determine the final leaderboard designs, 73\% of 15 testers preferred the \textit{hybrid-absolute} design for concepts. When asked to justify such a choice, participants claimed that in broader contexts like concepts, they primarily care about top users. They only want to see how top users learn and get inspired from them. Conversely, opposite results were found for courses. Indeed, 75\% chose the \textit{hybrid 50-50} solution. Here, users explained that since they have real relationships with other students, they also want to see similar classmates. This was justified for two reasons. Firstly, users want to assess themselves compared to the rest of the class. Secondly, users feel more comfortable to approach similar students rather than top ones \cite{mirkoThesis}. To validate leaderboards' effect on users' engagement, we performed a longitudinal validation. We collected users activities made through the leaderboards, for a total of 30 days. Even though the project constraints did not allow us to produce evidence with statistical significance, preliminary results suggest that users' activities are mainly focused on top positions. In fact, 55\% of the activities observed in the leaderboards occurred in the best users (1st position in the table). Moreover, more than 97\% happened within the first 4 positions.

\section{Conclusions and Future work}
\label{sec:conclusionsfuturework}

    FullBrain has shown a lot of promise to provide a digital home for learners. From the release of the social features on FullBrain (October 2020) until the current publication we have experienced an increase of $146\%$ in new registrations, totalling at approximately 500 users. In addition, we have described FullBrain's Social Information Retrieval System (\ac{SIR}) and leaderboards. The \ac{SIR} provides the following functions: Search, Query Autocomplete and Query Suggestions (no query input). These permit to explore platform's content, search through multiple entities while using the social graph to rank the results themselves. Moreover, the architecture is able to build the related user's responses in less than 0.11s. For the Leaderboards, we propose a hybrid-absolute design in concepts and a hybrid 50-50 design for courses, based on user interviews. Through preliminary analysis we recognise that Leaderboards produce user activity focused on the top ranking members with 97\% of activity directed to the top 4 positions.



    Larger user numbers allow us to experiment with Machine Learning and Deep Learning systems. Previous work by Filianos~\cite{panosThesis} has attempted to use Multigated Mixture of Expert (MMoE)~\cite{zhaommoeetal} and Mixture of Sequential Experts (MoSE)~\cite{zhenmsoe} models to rank posts in our home page stream. As our user base grows, we consider FullBrain to be very fertile ground for learning analytics and user activity dataset research.

\bibliographystyle{splncs04}
\bibliography{bibliography}

%




\end{document}